\documentclass[aps,pra,twocolumn,showpacs,amsmath]{revtex4-1}
\usepackage{amssymb,amsmath}
\usepackage{mathrsfs}
\usepackage{graphicx}
\usepackage{float}
\usepackage{txfonts}
\usepackage[position=t,singlelinecheck=off,
caption=false]{subfig}
\usepackage[normalem]{ulem}
\usepackage{color}
\usepackage{bm}
\usepackage{xcolor}
\usepackage{pdfcomment}
\usepackage{graphicx}
\usepackage{dcolumn}
\usepackage{bm}
\usepackage{bbm}
\usepackage{soul}
\setstcolor{red}

\usepackage[T1]{fontenc}
\usepackage[latin9]{inputenc}
\usepackage{setspace}
\usepackage{esint}

\begin{document}

\preprint{APS/123-QED}

\title{Floquet topological phases in a spin-$1/2$ double kicked rotor}

\author{Longwen \surname{Zhou}}\email{zhoulw13@u.nus.edu}
\affiliation{Department of Physics, National University of Singapore, Singapore 117551, Republic of Singapore}

\author{Jiangbin \surname{Gong}}\email{phygj@nus.edu.sg}
\affiliation{Department of Physics, National University of Singapore, Singapore 117551, Republic of Singapore}

\date{\today}

\begin{abstract}
The double kicked rotor model is  a physically realizable extension of the paradigmatic kicked rotor
model in the study of quantum chaos. Even before the concept of Floquet topological phases
became widely known, the discovery of
the Hofstadter butterfly spectrum in the double kicked rotor model {[}J. Wang and
J. Gong, Phys. Rev. A~\textbf{77}, 031405 (2008){]} already suggested the importance of periodic driving
to the generation of unconventional topological matter. In this work, we
explore Floquet topological phases of a double kicked rotor with an
\emph{extra} spin-$1/2$ degree of freedom. The latter has been experimentally
engineered in a quantum kicked rotor recently by loading $^{87}{\rm Rb}$
condensates into a periodically pulsed optical lattice. Under the
on-resonance condition, the spin-$1/2$ double kicked rotor admits
fruitful topological phases due to the interplay between its external
and internal degrees of freedom. Each of these topological phases
is characterized by a pair of winding numbers, whose combination
predicts the number of topologically protected $0$ and $\pi$-quasienergy
edge states in the system. Topological phases with arbitrarily large winding numbers can be easily found by tuning the kicking strength. We discuss an experimental proposal to realize this model
in kicked $^{87}{\rm Rb}$ condensates, and suggest
to detect its topological invariants by measuring the mean chiral
displacement in momentum space.
\end{abstract}

\maketitle

\section{Introduction}\label{sec:Int}

A topological characterization of a quantum chaos model by Leboeuf {\it et al}~\cite{LeboeufPRL1990} pioneered the use of periodic driving fields to create topological phases of matter absent in time-independent systems.  The model proposed in Ref.~\cite{LeboeufPRL1990} was however rather abstract because it is quantized on a phase space torus.  By extending the paradigmatic
kicked rotor model in the study of quantum chaos~\cite{KR1,KR2,KR3,KR4,KR5,KR6,OnRes1,OnRes2,KRRevs}, Wang and Gong proposed a physically realizable double kicked rotor model~\cite{DKRGong} and discovered Hofstadter's butterfly-like Floquet spectrum therein \cite{Butterfly}.  This finding strongly suggested that
such periodically driven systems are topologically rich and should be highly useful as dynamical systems to explore condensed-matter physics.  Indeed, the work by Wang and Gong~\cite{DKRGong}
has led to the proposal of a topological Thouless pump in momentum space \cite{DerekPRL2012}, the proof of the topological
equivalence between the double kicked rotor model and the kicked Harper model \cite{KHM,KHMDana,HailongPRE2013},
and the identification of many topological edge states in both of the two models \cite{DerekPRB2014}.

To date, Floquet topological states of matter have been well recognized as a promising concept and a fruitful topic. Floquet states, being intrinsically out-of-equilibrium,
can be engineered to carry topological properties that are either analogous to~\cite{OkaPRB2009,LindnerNP2011,DahlhausPRB2011,KitagawaPRB2011,TongPRB2013,LeonPRL2013,CayssolRRL2013,GrushinPRL2014,Wang2014,ZhouEPJB2014,TitumPRL2015,XiongPRB2016,DLossPRLPRB},
or even beyond their static cousins~\cite{KitagawaPRB2010,JiangPRL2011,KunduPRL2013,Bomantara2016,ZhaoErH2014,ReichlPRA2014,LeonPRB2014,AnomalousESPRX,FulgaPRB2016,YapPRB2017,YapMajorana2017,ZhouPRB2016}.
The latter includes, but is not limited to, degenerate $\pi$-quasienergy
edge states~\cite{JiangPRL2011,KunduPRL2013,Bomantara2016}, counterpropagating~\cite{ZhaoErH2014,ReichlPRA2014,LeonPRB2014} and anomalous chiral
edge states~\cite{AnomalousESPRX,FulgaPRB2016} in both insulating~\cite{YapPRB2017} and superconducting~\cite{YapMajorana2017} band
structures, leading to new types of topological classification schemes
and bulk-boundary relations~\cite{ZhouPRB2016,AsbothSTF,NathanNJP2015,ClassificationFTP1,ClassificationFTP2}. Accompanying great theoretical efforts in exploring these intriguing features~\cite{EckardtRMP2017}, Floquet topological states have also been
observed in several experimental settings, including ultracold
atom~\cite{ColdAtomFTP}, photonic~\cite{PhotonFTP0,PhotonFTP1,PhotonFTP2},
phononic and acoustic systems~\cite{PhononFTP}.

Motivated by recent experimental advances, here we continue to explore Floquet topological phases in the context of double kicked rotor model. The Hamiltonian of an earlier quantum double kicked rotor (DKR)~\cite{DKRGong} model, which was realized by cold atoms subjected to pairs of pulses in an optical lattice~\cite{DKREXP},
is given by
\begin{equation}
\hat{H}=\frac{\hat{p}^{2}}{2}+\kappa_{1}\cos(\hat{x})\sum_{m}\delta(t-mT)+\kappa_{2}\cos(\hat{x}+\beta)\sum_{m}\delta(t-mT-\tau).
\end{equation}
The stroboscopic dynamics of the system is governed by its evolution
operator over one $\delta$-kicking period $T$, {\it i.e.}, the Floquet
operator
\begin{equation}
\hat{F}=e^{-i(T-\tau)\frac{\hat{p}^{2}}{2\hbar}}e^{-i\frac{\kappa_{2}}{\hbar}\cos(\hat{x}+\beta)}e^{-i\tau\frac{\hat{p}^{2}}{2\hbar}}e^{-i\frac{\kappa_{1}}{\hbar}\cos(\hat{x})}. \end{equation}
Here all quantities are in dimensionless units. $\hat{x}$
and $\hat{p}$ are position and momentum operators for cold atoms.
$\beta$ is the phase shift between two kicking optical lattice potentials
of strengths $\kappa_{1}$ and $\kappa_{2}$, separated by a time
delay $\tau\in(0,T)$. Due to the spatial periodicity
of kicking potentials, the momentum $\hat{p}$ take values $p=(n+\eta)\hbar$,
where $\eta\in(0,1)$ is the conserved quasimomentum and $n\in\mathbb{Z}$.
For a Bose-Einstein condensate~(BEC) of large coherence width \cite{BetaEqT0,OnRes1}, $\eta$
can be set to zero, and $\hat{p}=\hat{n}\hbar$ only takes integer
multiples of Planck constant $\hbar$. Then under the on-resonance
condition~\cite{BetaEqT0,OnRes1,OnRes2} $\hbar T=4\pi$, the quantum
DKR has a Hofstadter's butterfly-like quasienergy spectrum~\cite{Butterfly},
characterized by fruitful topological band/gap structures and consecutive
topological phase transitions versus the change of the system's effective
Planck constant $\hbar$~\cite{DKRGong}.

In this work, we take one step further in the study of DKR by considering an internal spin-$1/2$ degree of freedom. The Floquet operator
of such a spin-$1/2$ double kicked rotor (DKRS) is given by
\begin{equation}
\hat{U}=e^{-i(T-\tau)\frac{\hat{p}^{2}}{2\hbar}}e^{-i\frac{\kappa_{2}}{\hbar}\cos(\hat{x}+\beta)\sigma_{y}}e^{-i\tau\frac{\hat{p}^{2}}{2\hbar}}e^{-i\frac{\kappa_{1}}{\hbar}\cos(\hat{x})\sigma_{x}},
\end{equation}
where $\sigma_{x,y,z}$ are Pauli matrices acting on internal spin
space of the rotor. More specifically, in the case of quasimomentum
$\eta=0$ \cite{BetaEqT0,OnRes1}, time delay $\tau=T/2$, and under
on-resonance condition~\cite{KRRevs,BetaEqT0} $\hbar\tau=4\pi$,
the Floquet propagator of DKRS reduces to
\begin{equation}
\hat{U}=e^{-iK_{2}\cos(\hat{x}+\beta)\sigma_{y}}e^{-iK_{1}\cos(\hat{x})\sigma_{x}},\label{eq:UORDKRS}
\end{equation}
where $K_{1,2}=\kappa_{1,2}/\hbar$ are rescaled kicking strengths.
In the following, we will first discuss a possible way of engineering
the on-resonance DKRS (ORDKRS) described by Eq.~(\ref{eq:UORDKRS}) in a periodically pulsed BEC, thanks to a recent experimental realization of quantum walks in momentum space~\cite{Wimberger,Wimberger2}. Next, we will explore the rich topological
phases of ORDKRS. Finally, we suggest to probe bulk topological invariants
of ORDKRS by measuring the mean chiral displacement of a wave packet
over tens of kicks, which is also experimentally available in both
photonic~\cite{MCD1} and cold atom~\cite{MCD2} systems.

\section{Realization of the ORDKRS}\label{sec:Realization}
The formalism of ORDKRS as described by Eq.~(\ref{eq:UORDKRS}) is
inspired by a recent experiment, which realizes discrete time quantum
walks in momentum space with a BEC of $^{87}{\rm Rb}$
\cite{Wimberger,Wimberger2}. The experimental platform is sketched
in Fig.~1 of Ref.~\cite{Wimberger}. Each step of the quantum walk
is composed of two consecutive operations. First, a resonant microwave
is applied to the $^{87}{\rm Rb}$ condensate, which introduces a
rotation within the two-state space of its ground hyperfine levels
$5^{2}S_{1/2}F=1$ and $5^{2}S_{1/2}F=2$. This realizes a ``coin
toss'' described by \cite{Wimberger,Wimberger2}
\begin{equation}
\mathscr{M}(\alpha,\chi)=e^{-i\frac{\alpha}{2}[\sin(\chi)\sigma_{x}-\cos(\chi)\sigma_{y}]},
\end{equation}
where $\sigma_{x,y,z}$ are Pauli matrices acting on the internal
two-state space, and the rotation angles $\alpha,\chi$ are controllable
experimentally. Next, the BEC is subjected to a short laser pulse,
whose frequency is detuned from the frequency between the two hyperfine
levels, realizing the far off-resonant condition and producing periodic
potentials. This step employs the atom-optical realization of the
quantum kicked rotor (ratchet accelerator) with a kicking strength
$k=\frac{\Omega^{2}\tau_{p}}{\Delta}$, where $\Omega$ is the Rabi
frequency, $\tau_{p}$ is the pulse length, and $\Delta$ is the detuning
of laser light from the atomic transition. Notably, the detuning $\Delta$
is positive for the state $5^{2}S_{1/2}F=1$ and negative for the
state $5^{2}S_{1/2}F=2$ of $^{87}{\rm Rb}$. Then under the quantum
on-resonance condition~ \cite{Wimberger,Wimberger2} (corresponding to the choice $\hbar\tau=4\pi$
in our model), the second operation in a quantum walk step is described
by a propagator~\cite{KRSOther}
\begin{equation}
\mathscr{T}=e^{-iK\cos(\hat{x})\sigma_{z}},
\end{equation}
where $K=\frac{\Omega^{2}\tau_{p}}{|\Delta|}$ is the absolute value
of kicking strength. The coupling between the internal degrees of
freedom (hyperfine levels $F=1,2$) and the external motion (hopping
in momentum space) is realized by the term $\cos(\hat{x})\sigma_{z}$.

The successful implementations of ``coin toss'' operation $\mathscr{M}(\alpha,\chi)$
and spin-dependent walk $\mathscr{T}$ in kicked BECs set the starting
point for the realization of an ORDKRS as described by Eq. (\ref{eq:UORDKRS}).
To see this, we rewrite the Floquet operator of ORDKRS as
\begin{equation}
\hat{U}=\hat{V}_{2}\hat{V}_{1},
\end{equation}
where $\hat{V}_{1}=e^{-iK_{1}\cos(\hat{x})\sigma_{x}}$ and $\hat{V}_{2}=e^{-iK_{2}\cos(\hat{x}+\beta)\sigma_{y}}$.
Then each of these two propagators can be realized by proper combinations
of ``coin toss'' and spin-dependent walk operations:
\begin{alignat}{1}
\hat{V}_{1}= & e^{-iK_{1}\cos(\hat{x})\sigma_{x}}=\mathscr{M}\left(-\frac{\pi}{2},0\right)\mathscr{T}_{1}\mathscr{M}\left(\frac{\pi}{2},0\right),\\
\hat{V}_{2}= & e^{-iK_{2}\cos(\hat{x}+\beta)\sigma_{y}}=\mathscr{M}\left(-\frac{\pi}{2},\frac{\pi}{2}\right)\mathscr{T}_{2}\mathscr{M}\left(\frac{\pi}{2},\frac{\pi}{2}\right),
\end{alignat}
where $\mathscr{T}_{1}=e^{-iK_{1}\cos(\hat{x})\sigma_{z}}$ and $\mathscr{T}_{2}=e^{-iK_{2}\cos(\hat{x}+\beta)\sigma_{z}}$
are two spin-dependent walks. The different kicking strengths $K_{1,2}=\Omega^2_{1,2}\tau_p/|\Delta_{1,2}|$
may be realized by letting the two walks to have either a different
Rabi frequency $\Omega_1\neq\Omega_2$ or a different detuning $|\Delta_1|\neq|\Delta_2|$. Putting
together, the Floquet operator of ORDKRS is realized by a sequence
of operations as $\hat{U}=\mathscr{M}\left(-\frac{\pi}{2},\frac{\pi}{2}\right)\mathscr{T}_{2}\mathscr{M}\left(\frac{\pi}{2},\frac{\pi}{2}\right)\mathscr{M}\left(-\frac{\pi}{2},0\right)\mathscr{T}_{1}\mathscr{M}\left(\frac{\pi}{2},0\right)$.
Since each sub-step in this sequence is already realized in the quantum
walk experiment of $^{87}{\rm Rb}$ condensates \cite{Wimberger,Wimberger2},
the realization of ORDKRS as described by Eq. (\ref{eq:UORDKRS})
should already be available in the same experimental setup or other
similar platforms.

To further motivate experimental interests, we will analyze the topological
properties of ORDKRS in the following sections. To be more explicit,
we choose the phase shift between the two kicks to be $\beta=-\frac{\pi}{2}$
in Eq. (\ref{eq:UORDKRS}). This gives us the following Floquet operator
of an ORDKRS:
\begin{equation}
\hat{U}_{R}=e^{-iK_{2}\sin(\hat{x})\sigma_{y}}e^{-iK_{1}\cos(\hat{x})\sigma_{x}}.\label{eq:UR}
\end{equation}
As will be shown, this system possesses a fruitful Floquet topological
phases, with their topological winding numbers detectable by measuring
momentum distributions of the system over tens of driving periods.

Note in passing that by choosing the initial state to be a coherent superposition of several momentum eigenstates~\cite{Wimberger,Wimberger2}, the Floquet operator $\hat{U}_{R}$ may also be used to engineer a split step quantum walk in the momentum space of BECs, whose topological properties have been thoroughly explored in previous studies~\cite{QWReview}. Compared with the split step quantum walk, the ORDKRS introduced here admits a richer topological phase diagram, with the possibility to access phases with large topological invariants.

\section{Topological phases of the ORDKRS}\label{sec:FTPs}
Similar to their static cousins \cite{Tenfold}, single-particle Floquet
topological phases in one-dimension are all symmetry protected \cite{ClassificationFTP1}.
The Floquet operator $\hat{U}_{R}$, as defined in Eq.~(\ref{eq:UR}),
possesses a chiral symmetry. Its topological phases are then characterized
by a pair of integers ($\mathbb{Z}\times\mathbb{Z}$), defined in
two complementary chiral symmetric time frames~\cite{AsbothSTF}.
These integers predict the number of degenerate $0$ and $\pi$-quasienergy
edge states in the two spectrum gaps of $\hat{U}_{R}$, respectively.
These will be demonstrated in the following subsections.

\subsection{Chiral symmetric time frame and topological winding number}\label{sec:WN}
The chiral symmetry of $\hat{U}_{R}$ is most clearly seen by transforming
it into two chiral symmetric time frames \cite{AsbothSTF}, in which
it has the following forms:
\begin{alignat}{1}
\hat{U}_{1}= & e^{-i\frac{K_{1}}{2}\cos(\hat{x})\sigma_{x}}e^{-iK_{2}\sin(\hat{x})\sigma_{y}}e^{-i\frac{K_{1}}{2}\cos(\hat{x})\sigma_{x}},\label{eq:UFrame1}\\
\hat{U}_{2}= & e^{-i\frac{K_{2}}{2}\sin(\hat{x})\sigma_{y}}e^{-iK_{1}\cos(\hat{x})\sigma_{x}}e^{-i\frac{K_{2}}{2}\sin(\hat{x})\sigma_{y}}.\label{eq:UFrame2}
\end{alignat}
It is seen that both $\hat{U}_{1}$ and $\hat{U}_{2}$ are related
to $\hat{U}_{R}$ by unitary transformations, meaning that they all
share the same Floquet quasienergy spectrum. Furthermore, both $\hat{U}_{1}$
and $\hat{U}_{2}$ possess the chiral symmetry as
\begin{equation}
\Gamma\hat{U}_{1}\Gamma=\hat{U}_{1}^{\dagger},\qquad\Gamma\hat{U}_{2}\Gamma=\hat{U}_{2}^{\dagger},\qquad\Gamma=\sigma_{z}.
\end{equation}
Here the chiral symmetry operator $\Gamma$ is both Hermitian and
unitary, {\it i.e.}, $\Gamma=\Gamma^{\dagger}=\Gamma^{-1}$. Based on the
periodic table of Floquet topological states~\cite{ClassificationFTP1},
each phase of $\hat{U}_{R}$ is then characterized by a pair of integer
winding numbers $\left(W_{0},W_{\pi}\right)\in\mathbb{Z}\times\mathbb{Z}$
\cite{AsbothSTF}, given by
\begin{equation}
W_{0}=\frac{W_{1}+W_{2}}{2},\qquad W_{\pi}=\frac{W_{1}-W_{2}}{2},\label{eq:WN}
\end{equation}
where $W_{1}$ and $W_{2}$ are winding numbers of Floquet operators
$\hat{U}_{1}$ and $\hat{U}_{2}$, respectively. The winding numbers $\left(W_{0},W_{\pi}\right)$ allow us to achieve a full classification of the topological phases of $\hat{U}_R$, as will be discussed in Sec.~\ref{sec:PhsDiagram}.

To compute these winding numbers for each Floquet topological phase,
we rewrite $\hat{U}_{\ell}$ ($\ell=1,2$) by combining its three
pieces. In the position representation $\{|\theta\rangle|\theta\in[-\pi,\pi)\}$,
we then have $\hat{U}_{\ell}=\sum_{\theta}|\theta\rangle\langle\theta|e^{-iE(\theta){\bf n}_{\ell}\cdot\boldsymbol{\sigma}}$.
The dispersion $E(\theta)$ has the form (see Appendix~\ref{app:Ul} for more details)
\begin{equation}
E(\theta)=\arccos[\cos({\cal K}_{1})\cos({\cal K}_{2})],\label{eq:ETheta}
\end{equation}
where ${\cal K}_{1}\equiv K_{1}\cos\theta$ and ${\cal K}_{2}\equiv K_{2}\sin\theta$.
The vector of matrix $\boldsymbol{\sigma}=(\sigma_{x},\sigma_{y})$,
and the two-component unit vectors ${\bf n}_{\ell}=(n_{\ell x},n_{\ell y})$
for $\ell=1,2$ are explicitly given by
\begin{alignat}{1}
n_{1x}= & \frac{\sin({\cal K}_{1})\cos({\cal K}_{2})}{\sqrt{\sin^{2}({\cal K}_{1})\cos^{2}({\cal K}_{2})+\sin^{2}({\cal K}_{2})}},\label{eq:N1X}\\
n_{1y}= & \frac{\sin({\cal K}_{2})}{\sqrt{\sin^{2}({\cal K}_{1})\cos^{2}({\cal K}_{2})+\sin^{2}({\cal K}_{2})}},\label{eq:N1Y}
\end{alignat}
and
\begin{alignat}{1}
n_{2x}= & \frac{\sin({\cal K}_{1})}{\sqrt{\sin^{2}({\cal K}_{1})\cos^{2}({\cal K}_{2})+\sin^{2}({\cal K}_{2})}},\label{eq:N2X}\\
n_{2y}= & \frac{\sin({\cal K}_{2})\cos({\cal K}_{1})}{\sqrt{\sin^{2}({\cal K}_{1})\cos^{2}({\cal K}_{2})+\sin^{2}({\cal K}_{2})}},\label{eq:N2Y}
\end{alignat}
Using these vectors, the winding number $W_{\ell}$ of Floquet operator
$\hat{U}_{\ell}$ \cite{MCD1} can be computed as
\begin{equation}
W_{\ell}=\int_{-\pi}^{\pi}\frac{d\theta}{2\pi}({\bf n}_{\ell}\times\partial_{\theta}{\bf n}_{\ell})_{z},\qquad\ell=1,2.\label{eq:WN12}
\end{equation}
As evidenced by this expression, the winding number $W_{\ell}$ counts
the number of times that the unit vector ${\bf n}_{\ell}$ rotates around
the $z$-axis when $\theta$ changes over a period from $-\pi$ to
$\pi$. Thanks to the chiral symmetry of $\hat{U}_{\ell}$, the vector
${\bf n}_{\ell}$ is constrained to rotate on the $x$-$y$ plane,
ensuring $W_{\ell}$ to be a well defined integer. Furthermore, the
quantization of the winding number $W_{\ell}$ is topologically protected,
since $W_{\ell}$ cannot change its value under continuous deformations
of the trajectory of ${\bf n}_{\ell}$ on the $x$-$y$ plane. The
topological property of winding numbers $\left(W_{0},W_{\pi}\right)$
are then carried over from winding numbers $W_{1}$ and $W_{2}$ through
Eq.~(\ref{eq:WN}).

\subsection{Topological phase diagram}\label{sec:PhsDiagram}
If the trajectory of vector ${\bf n}_{\ell}$ on the $x$-$y$ plane
happens to pass through the origin of $z$-axis at some critical value $\theta=\theta_{c}$,
the dispersion $E(\theta)$ will become gapless. This situation indicates
the breakdown of the winding number definition (\ref{eq:WN12}) and the
existence of a possible topological phase transition specified by
its corresponding kicking strengths $(K_{1c},K_{2c})$. The collection
of all these transition points on the plane of parameter space $(K_{1},K_{2})$
forms the boundary between different Floquet topological phases of
the ORDKRS.

To locate these phase boundaries, we note that being a phase factor defined
modulus $2\pi$, the dispersion $E(\theta)$ has in general two gaps
at both quasienergies $0$ and $\pi$, respectively. The closure of
a spectrum gap in $E(\theta)$ then corresponds to either $E(\theta)=0$
or $E(\theta)=\pi$, which means that $\cos({\cal K}_{1})\cos({\cal K}_{2})=\pm1$
in Eq. (\ref{eq:ETheta}), respectively. This condition can be met
if and only if $K_{1}\cos\theta=\mu\pi$ and $K_{2}\sin\theta=\nu\pi$, where
$\nu,\mu$ are both integers. The combination of these conditions
yields the following equation for the topological phase boundaries
of $\hat{U}_{R}$:
\begin{equation}
\frac{\mu^{2}}{K_{1}^{2}}+\frac{\nu^{2}}{K_{2}^{2}}=\frac{1}{\pi^{2}},\qquad\mu,\nu\in\mathbb{Z}.
\end{equation}
Following their experimental definitions, we focus on the regime of
positive kicking strengths $K_{1},K_{2}>0$. In this regime, the phase
boundaries can be classified into three groups based on the value
of integers $(\mu,\nu)$.

(i) $\mu=0$: In this case, the phase boundaries $K_{2}=\nu\pi$ ($\nu=0,1,2,...$)
are straight lines in parallel with the $K_{1}$-axis on the $K_{1}$-$K_{2}$
plane. Furthermore, when $\nu$ is an odd (even) integer, the Floquet
spectrum gap will close at quasienergy $\pi$ ($0$). The corresponding
topological phase transition is only accompanied by the change of
winding number $W_{\pi}\,(W_{0})$.

(ii) $\nu=0$: In this case, the phase boundaries $K_{1}=\mu\pi$
($\mu=0,1,2,...$) are straight lines in parallel with the $K_{2}$-axis
on the $K_{1}$-$K_{2}$ plane. Furthermore, when $\mu$ is an odd
(even) integer, the Floquet spectrum gap will close at quasienergy
$\pi$ ($0$). The corresponding topological phase transition is only
accompanied by the change of winding number $W_{\pi}\,(W_{0})$.

(iii) $\mu,\nu\neq0$: In this case, the phase boundary curves are
described by the equation $\frac{K_{2}}{\pi}=\nu\left(1-\frac{\mu^{2}}{K_{1}^{2}/\pi^{2}}\right)^{-1/2}$,
with positive solutions only for $K_{1}>\mu\pi$. Furthermore, when $\mu,\nu$
have the opposite (same) parities, the Floquet spectrum gap will close
at quasienergy $\pi$ ($0$) along the phase boundary curve. The corresponding topological phase
transition is only accompanied by the change of winding number $W_{\pi}\,(W_{0})$.
\begin{figure}
	\centering
	\includegraphics[width=.47\textwidth]{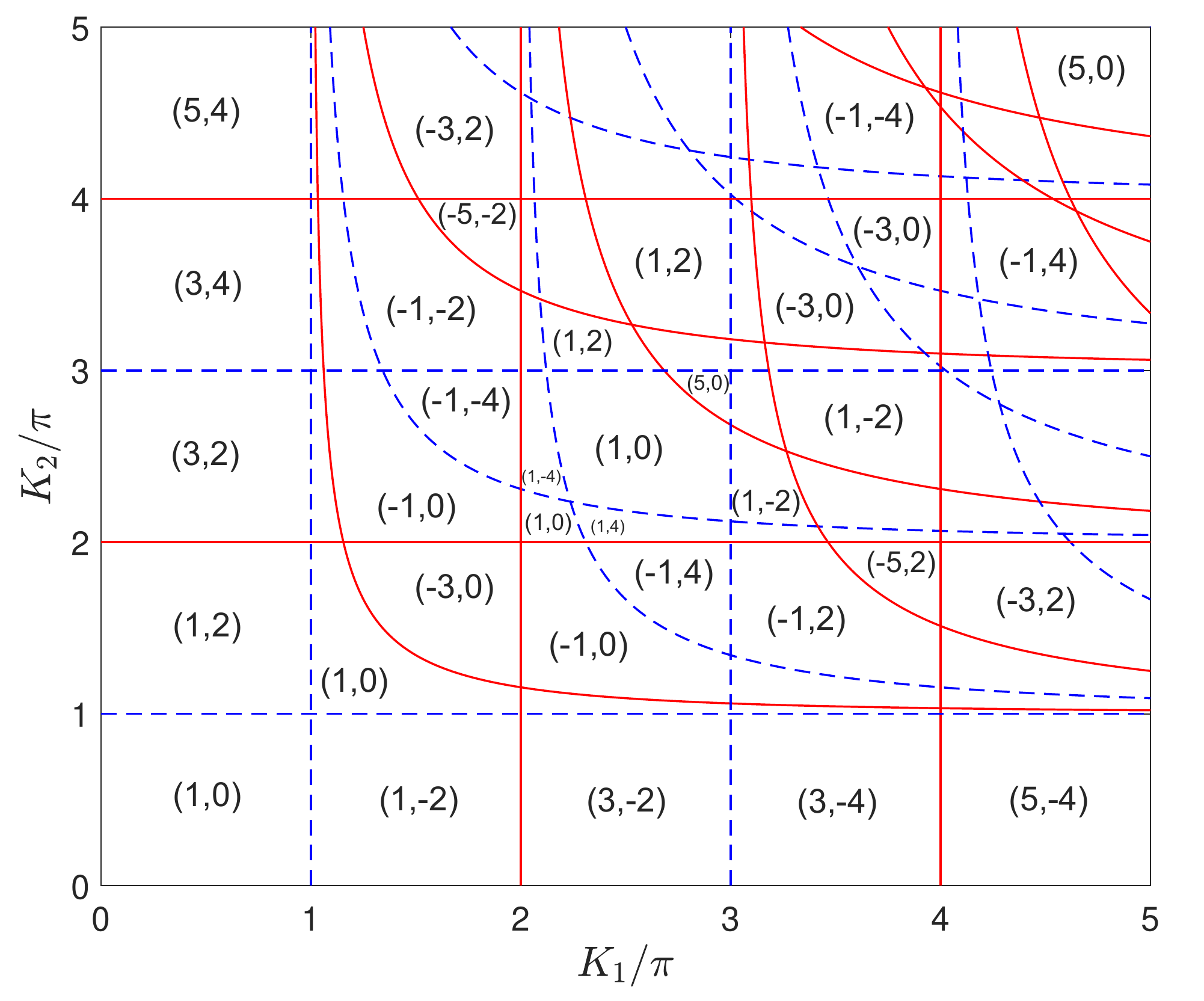}
	\caption{(color online) Floquet topological phase diagram of the ORDKRS $\hat{U}_R$ versus kicking strengths
		$(K_{1},K_{2})$. Red solid (blue dashed) lines are phase boundaries,
		where the Floquet spectrum gap close at quasienergy $0$ ($\pi$).
		Each closed patch corresponds to a unique topological phase, characterized
		by a pair of winding numbers $(W_{0},W_{\pi})$ deduced from Eq. (\ref{eq:WN}),
		as denoted in the figure for some representative phases.}
	\label{fig:PhsDiagram}
\end{figure}

Combining points (i-iii) together with winding numbers calculated
from Eq.~(\ref{eq:WN}), we are able to achieve a full topological
classification of the ORDKRS as described by the Floquet operator
$\hat{U}_{R}$ in Eq.~(\ref{eq:UR}). A topological phase diagram
of the system up to $K_{1}=K_{2}=5\pi$ is shown in Fig.~\ref{fig:PhsDiagram}. On the
phase diagram, each closed patch is characterized by a pair of winding
numbers $(W_{0},W_{\pi})$.

In Ref.~\cite{DerekPRB2014}, a phase diagram with similar phase boundaries is found in a spinless DKR model under a different on-resonance condition. Notably, the topological phase in each patch of that phase diagram is characterized by different winding numbers from that of the ORDKRS studied here. This difference comes from distinct winding behaviors of the vector ${\bf n}_\ell$ in the two models, even though they share the same Floquet spectrum.

Furthermore, in the region $(K_1,K_2)\in(0,\pi)\times(0,\infty)$ ($(K_1,K_2)\in(0,\infty)\times(0,\pi)$), the winding numbers $W_0$ and $W_{\pi}$ both tend to grow linearly along the direction of $K_2$- ($K_1$-)axis without bound (see Appendix.~\ref{app:WNgrowth} for an illustration). This result mimics the change of quantum Hall resistance (here the winding number) with the increase of a magnetic field (here the kicking strength) in quantum Hall effects~\cite{QHEplateaus,KR4}. A similar pattern is also observed in the phase diagram of the spinless DKR studied in Ref.~\cite{DerekPRB2014}. The possibility of accessing phases with arbitrarily large winding numbers in the ORDKRS makes it a good candidate to explore Floquet states and phase transitions in the regime of large topological invariants, which is usually absent in other experimentally realized one-dimensional Floquet systems like the split step quantum walk~\cite{QWReview}.

In the next subsection, we will explore the relation between the winding numbers of $\hat{U}_{R}$ and the number of its topological edge states in a finite-size momentum space lattice.

\subsection{Bulk-boundary correspondence}\label{sec:BBC}
The Floquet operator $\hat{U}_{R}=e^{-iK_{2}\sin(\hat{x})\sigma_{y}}e^{-iK_{1}\cos(\hat{x})\sigma_{x}}$
can be written in momentum representation~\cite{DerekPRB2014} as
\begin{equation}
\hat{U}_{R}=e^{-iK_{2}\sum_{n}\frac{1}{2i}(|n\rangle\langle n+1|-{\rm h.c.})\sigma_{y}}e^{-iK_{1}\sum_{n}\frac{1}{2}(|n\rangle\langle n+1|+{\rm h.c.})\sigma_{x}},\label{eq:URMom}
\end{equation}
where the momentum basis $\{|n\rangle|n\in\mathbb{Z}\}$ satisfies the
eigenvalue equation $\hat{n}|n\rangle=n|n\rangle$, with $\hat{n}$
being the dimensionless momentum operator as discussed in Sec.~\ref{sec:Int}.
This result can be obtained, {\it e.g.}, by first writing $K_{1}\cos(\hat{x})\sigma_{x}$
in position representation as $K_{1}\sum_{\theta}\frac{e^{i\theta}+e^{-i\theta}}{2}|\theta\rangle\langle\theta|\sigma_{x}$,
and then performing a Fourier transform from position to momentum
representation as $|\theta\rangle=\frac{1}{\sqrt{N}}\sum_{n=-\frac{N}{2}}^{\frac{N}{2}-1}e^{in\theta}|n\rangle$ under the periodic boundary condition $|n\rangle=|n+N\rangle$.
Expressed in the form of Eq.~(\ref{eq:URMom}), $\hat{U}_{R}$ admits an interpretation of
two consecutive kicks by momentum space tight-binding lattices on
a spin-$1/2$ particle. If an open boundary condition can be introduced
into this momentum space lattice, there will be topological edge states
localized around its boundaries if the kicking strengths $(K_{1},K_{2})$
reside in a topologically nontrivial patch of the phase diagram. This
is guaranteed by the bulk-boundary correspondence of chiral symmetric
Floquet systems \cite{AsbothSTF}. More precisely, the absolute value
of winding number $W_{0}$ ($W_{\pi}$) gives the number of degenerate
edge state pairs at quasienergy $0$ ($\pi$) in the momentum space
lattice.
\begin{figure}
	\centering
	\includegraphics[width=.48\textwidth]{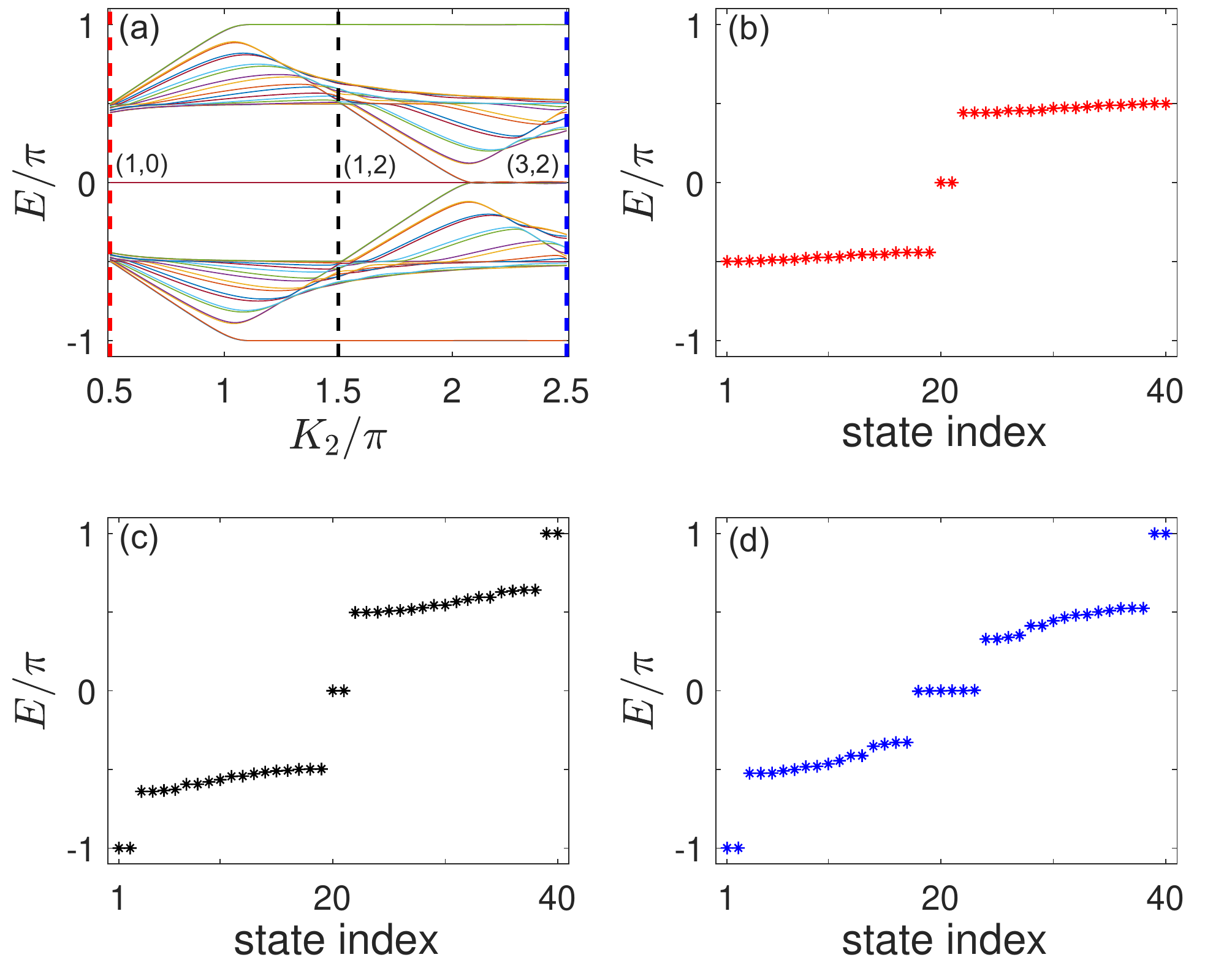}
	\caption{(color online) Bulk-boundary correspondence of the ORDKRS. Panel (a): Floquet spectrum
		$E$ of $\hat{U}_{R}$ versus $K_{2}$ at $K_{1}=0.5\pi$, for a momentum
		space lattice of $N=20$ unit cells under open boundary conditions.
		Three topological phases with winding numbers $(W_{0},W_{\pi})=(1,0)$,
		$(1,2)$ and $(3,2)$, as denoted on the figure, are separated by
		two transitions at $K_{2}=\pi,2\pi$. Panel (b): Floquet spectrum
		$E$ at $K_{2}=0.5\pi$, referring to the cut along the red dashed
		line at the left end of Panel (a). There is a pair of $0$-quasienergy
		edge states, corresponding to $W_{0}=1$. Panel (c): Floquet spectrum
		$E$ at $K_{2}=1.5\pi$, referring to the cut along the black dashed
		line in the middle of Panel (a). There is a pair of $0$- and two
		pairs of $\pi$-quasienergy edge states, corresponding to $W_{0}=1$
		and $W_{\pi}=2$. Panel (d): Floquet spectrum $E$ at $K_{2}=2.5\pi$,
		referring to the cut along the blue dashed line at the right end of
		Panel (a). There are three pairs of $0$- and two pairs of $\pi$-quasienergy
		edge states, corresponding to $W_{0}=3$ and $W_{\pi}=2$.}
	\label{fig:BBC}
\end{figure}

An illustration of this bulk-boundary relation is given in Fig.~\ref{fig:BBC}.
The panel (a) of Fig.~\ref{fig:BBC} shows the spectrum of $\hat{U}_{R}$ at a fixed
value of the first kicking strength $K_{1}=0.5\pi$ under open boundary
conditions. With the change of the second kicking strength $K_{2}$,
the system undergoes two topological phase transitions, with quasienergy
gap closing at $\pi$ ($0$) for $K_{2}=\pi$ ($K_{2}=2\pi$). These
transitions separate the system in the considered range of parameters
into three different topological phases, characterized by winding
numbers $(W_{0},W_{\pi})=(1,0)$, $(1,2)$ and $(3,2)$ (See also
Fig.~\ref{fig:PhsDiagram}). These numbers correctly predict the number of $0$- and $\pi$-quasienergy
edge state pairs in these three topological phases, as exemplified by
panels (b) to (d) of Fig.~\ref{fig:BBC}. On the other hand, by counting the number
of $0$ and $\pi$ edge state pairs in Fig.~\ref{fig:BBC}(b-d), we can also obtain
the winding numbers $(W_{0},W_{\pi})$ for each topological phases.
This concludes the verification of bulk boundary correspondence in
the chiral symmetric ORDKRS system.

As a notable feature of Fig.~\ref{fig:BBC}(a), there are regions in which the $0$ and $\pi$ quasienergy edge states coexist at the same system parameters [see Fig.~\ref{fig:BBC}(c) or \ref{fig:BBC}(d) as an example]. In a recent study~\cite{RadityaMTC}, it was shown that a superposition of $0$ and $\pi$ edge states form a new type of symmetry protected discrete time crystal phase, which is further used to propose a new approach to non-Abelian braiding and topological quantum computing in a superconducting Floquet system. The unbounded growth of winding numbers in the phase diagram Fig.~\ref{fig:PhsDiagram} then implies the possibility of finding an arbitrarily large number of $0$ and $\pi$ quasienergy edge states at the same parameter of the ORDKRS, and therefore the potential of engineering many different Floquet time crystal phases~\cite{TCReview} in this system by superposing these edge states.

Experimentally, Floquet edge states between systems with different
bulk topological properties have been observed in photonic quantum
walks \cite{PhotonFTP0}. However, for the ORDKRS defined in a momentum
lattice as Eq. (\ref{eq:URMom}), it may not be easy to engineer a
boundary between different momentum space regions. In the following
section, we discuss an alternative way of detecting topological winding
numbers of the ORDKRS by directly imaging the momentum distribution
of a wave packet \cite{MCD2}, which is available in kicked BEC experimental
setups \cite{Wimberger}.

\section{Probing bulk topological properties of the ORDKRS}\label{sec:MCD}
The topological winding numbers of a one-dimensional chiral symmetric
system can be detected by measuring the mean chiral displacement (MCD)
of a wave packet \cite{MCD1,MCD2}. Formally, it is the expectation
value of chiral displacement operator $\hat{C}(t)=\hat{U}^{\dagger}(t)\hat{n}\Gamma\hat{U}(t)$
at some time $t$ of the system's unitary evolution $\hat{U}(t)$.
For the ORDKRS, $\hat{n}$ and $\Gamma=\sigma_{z}$ represent the
quantized momentum and chiral symmetry operators, respectively. Therefore
the MCD of ORDKRS is just a signed momentum distribution, with the
extra sign originating from the chiral symmetry. For the system of
$^{87}{\rm Rb}$ BECs prepared in the state $|\psi_{0}\rangle=|n=0,5^{2}S_{1/2}F=1\rangle$
or $|n=0,5^{2}S_{1/2}F=2\rangle$ of the $n=0$-momentum sector at
time $t=0$, the MCD after $t$ driving periods reads
\begin{equation}
C_{\ell}(t)=\langle\psi_{0}|\hat{U}_{\ell}^{-t}(\hat{n}\otimes\sigma_{z})\hat{U}_{\ell}^{t}|\psi_{0}\rangle,
\end{equation}
where the Floquet operators $\hat{U}_{\ell}$ ($\ell=1,2$) are given
by Eqs.~(\ref{eq:UFrame1}) and (\ref{eq:UFrame2}). Further calculations
lead to (see Appendix~\ref{app:MCD} for details):
\begin{equation}
C_{\ell}(t)=\frac{W_{\ell}}{2}-\int_{-\pi}^{\pi}\frac{d\theta}{2\pi}\frac{\cos[E(\theta)t]}{2}\left({\bf n}_{\ell}\times\partial_{\theta}{\bf n}_{\ell}\right)_{z}\qquad\ell=1,2.\label{eq:MCD}
\end{equation}
Here $W_{\ell}$ is the winding number of $\hat{U}_{\ell}$ given
by Eq. (\ref{eq:WN12}). The dispersion $E(\theta)$ is given by Eq.
(\ref{eq:ETheta}), and the components of unit vector ${\bf n}_{\ell}$
are given by Eqs. (\ref{eq:N1X}-\ref{eq:N2Y}). As can be seen, $C_{\ell}(t)$
contains a time-independent topological part $\frac{W_{\ell}}{2}$
and an extra time-dependent oscillating term. For a not-too-flat dispersion
$E(\theta)$, the oscillating term will tend to vanish for large $t$ under the
integral over $\theta$. A bit more rigorously, the $C_{\ell}(t)$
averaged over $t$ driving periods, {\it i.e.},
\begin{alignat}{1}
\overline{C_{\ell}(t)}
\equiv & \frac{1}{t}\sum_{t'=1}^{t}C_{\ell}(t')\nonumber\\
     = & \frac{W_{\ell}}{2}-\frac{1}{t}\sum_{t'=1}^{t}\int_{-\pi}^{\pi}\frac{d\theta}{2\pi}\frac{\cos[E(\theta)t]}{2}\left({\bf n}_{\ell}\times\partial_{\theta}{\bf n}_{\ell}\right)_{z},\label{eq:TAMCD}
\end{alignat}
will gradually converge to half of the winding number $W_{\ell}$
with the increase of $t$. Once $\frac{W_{1}}{2}$ and $\frac{W_{2}}{2}$
are obtained from the time averaged MCD, the winding numbers $(W_{0},W_{\pi})$
characterizing topological phases of the ORDKRS can be calculated
by Eq. (\ref{eq:WN}).
\begin{figure}
	\centering
	\includegraphics[width=.45\textwidth]{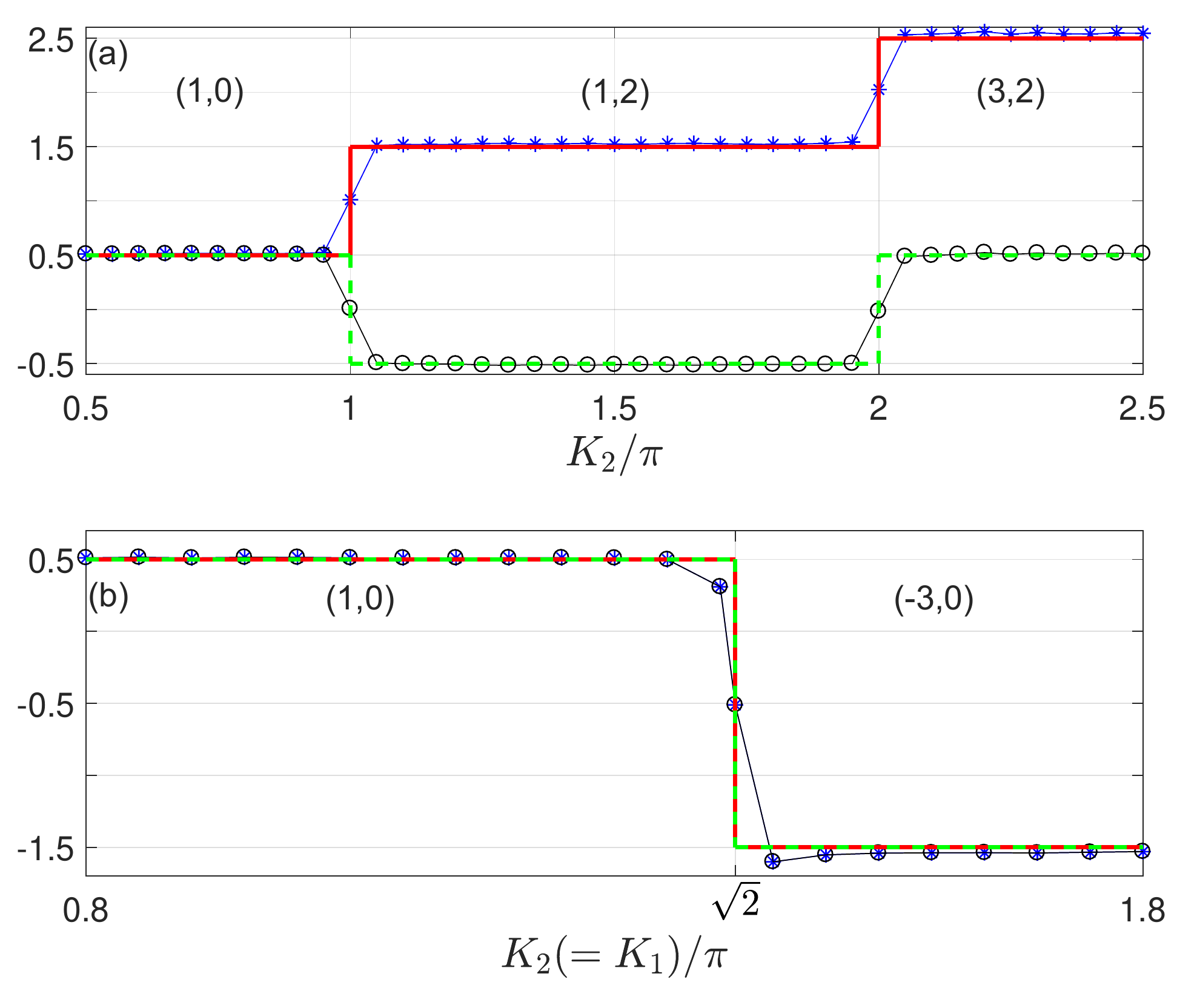}
	\caption{(color online) Time averaged MCDs. Numerical values of $\overline{C_{1}(t)}$ and
		$\overline{C_{2}(t)}$, both averaged over $t=20$ driving periods,
		are shown by blue stars and black circles. Theoretical values of half
		winding numbers $\frac{W_{1}}{2}$ and $\frac{W_{2}}{2}$ are shown
		by red solid and green dashed lines. In Panel (a), the kicking strength
		$K_{1}=0.5\pi$ and the three topological phases, separated by two
		transitions at $K_{2}=\pi,2\pi$, have winding numbers $(W_{0},W_{\pi})=\left(\frac{W_{1}+W_{2}}{2},\frac{W_{1}-W_{2}}{2}\right)=(1,0)$,
		$(1,2)$ and $(3,2)$, as denoted in the figure. In Panel (b), the
		two topological phases separated by a transition at $K_{1}=K_{2}=\sqrt{2}\pi$
		have winding numbers $(W_{0},W_{\pi})=\left(\frac{W_{1}+W_{2}}{2},\frac{W_{1}-W_{2}}{2}\right)=(1,0)$
		and $(-3,0)$, as denoted in the figure.}
	\label{fig:MCD}
\end{figure}

In Fig.~\ref{fig:MCD}, we present numerical results of $\overline{C_{\ell}(t)}$
along two different trajectories in the $K_{1}$-$K_{2}$ parameter
space, together with theoretical values of half winding numbers $W_1/2$ and $W_2/2$. The numerical results at each set of kicking strengths $(K_{1},K_{2})$ are obtained
by directly evolving a wave packet, prepared at initial state $|n=0,F=1\rangle$
or $|n=0,F=2\rangle$, with propagators $\hat{U}_{1}^{t}$ and $\hat{U}_{2}^{t}$
in momentum space to find $C_{1}(t)$ and $C_{2}(t)$, respectively,
and then averaging over the number of driving periods $t$. Up to
$t=20$, we find already very nice convergence of $\overline{C_{\ell}(t)}$
($\ell=1,2$) to its corresponding half winding number $\frac{W_{\ell}}{2}$,
with the error accounted for by the time-dependent term in Eq.~(\ref{eq:TAMCD}).
In the setup of $^{87}{\rm Rb}$ BEC, an implementation of
up to $50$ kicks is mentioned to be experimentally available~\cite{Wimberger}.
This corresponds to $t=25$ driving periods in our double kicked rotor,
more then the number needed to see a nice convergence in our numerical
simulations.

Recently, the measurements of MCD have been achieved in both photonic~\cite{MCD1} and cold atom~\cite{MCD2}
systems. In Ref.~\cite{MCD1}, the MCD is extracted from a quantum walk of twisted photons over $7$ steps, and the measured results are robust to dynamical disorder. In Ref.~\cite{MCD2}, $^{87}$Rb condensates are illuminated by a pair of off-resonant lasers to realize a synthetic lattice in momentum space. The coupling between adjacent momentum sites in this setup is controlled by two-photon Bragg transitions, and can be periodically quenched in time. In the high-frequency driving regime, the effective tight-binding Hamiltonian of the system falls into AIII or BDI topological class~\cite{Tenfold}. Disorder-induced topological phase transitions are then detected by meansing the MCD. Based on these facts, we believe that the
realization of ORDKRS and measurements of its topological winding
numbers are readily doable under current experimental conditions.

\section{Conclusions}\label{sec:Conclusions}
In this work, we proposed a spin-$1/2$ on-resonance double kicked
rotor model, which is realizable in BECs of ${^{87}}{\rm Rb}$ subjected
to pairs of periodic pulses by an optical lattice. The system owns
many intriguing Floquet topological phases, each characterized by
a pair of winding numbers and protected by the chiral symmetry of
the Floquet operator. Using these winding numbers, a full topological
phase diagram of the system was established. Under open boundary conditions,
this pair of winding numbers could also predict the number of topologically
protected edge state pairs at $0$ and $\pi$-quasienergies of the
Floquet spectrum. Finally, we proposed to detect these topological
winding numbers by measuring the mean chiral displacement of a wave
packet, initially localized at the center of the momentum space. The
numerical values of mean chiral displacement, averaged over $20$
kicking periods, tend to converge to the theoretical prediction
of bulk winding numbers of the ORDKRS. Recently, the experimental
measurements of mean chiral displacements have also been achieved in
other model systems~\cite{MCD1,MCD2}.

Our choice of the on-resonance condition, {\it i.e.}, $\hbar\tau=4\pi$ with
$\tau=T/2$, makes the free evolution part of the Floquet operator to become an identity.
Under more general resonance conditions, the free evolution
part can also contribute to the dynamics. The resulting Floquet operators
could then possess more then two Floquet bands and different types
of topological phases, as already indicated in a previous study of
the spinless quantum DKR~\cite{DerekPRB2014}. Exploring
the impact of an extra spin degree of freedom on the topological phases
of the DKR under general resonance conditions is an interesting topic for future study.

Due to experimental constrains on the detection window of momentum
states, only small to intermediate values of kicking strength are
considered in our numerical simulations. When the kicking strength
is large, the dynamics of the spin-$1/2$ double kicked rotor will
in general become chaotic in its classical limit. Exploring quantum
dynamics and its possible topological signatures in this classically
chaotic regime is certainly an intriguing topic. A recent study found
that up to large enough kicking strengths, the winding numbers $W$
of a periodically quenched chiral symmetric Floquet system satisfy
a Gaussian distribution around $W=0$ \cite{WNFluct2018}. Initial
numerical calculations in our system suggest a similar pattern along
the line $K_{2}=\lambda K_{1}$ on the phase diagram for any $|\lambda|\in(0,\infty)$.
However, for trajectories parallel to $K_{1}$ or $K_{2}$ axis on
the phase diagram and constrained within $K_{2}\in(0,\pi)$ or $K_{1}\in(0,\pi)$
regions, respectively, the winding numbers $W$ change monotonically
with the kicking strength and satisfy instead a uniform distribution.
The qualitative difference between these two types of winding number
distributions, the transition between them, and its possible connection
to the quantum-to-classical transitions in ORDKRS also deserve further explorations.

Finally, the effect of disorder on Floquet topological phases is of great theoretical and experimental interests \cite{TitumPRL2015,AnomalousESPRX,MCD1,MCD2}.
In a chiral symmetric system realized by quantum walk of twisted photons,
the Floquet topological phases have been demonstrated to be robust
to weak temporal disorder \cite{MCD1}. Furthermore, disorder induced
transitions from topological Anderson insulator to normal insulator
phases, and even the reverse, have also been observed quite recently
in the momentum space of laser driven ultracold atoms \cite{MCD2}.
One limitation of the models explored in these experiments is that
their winding numbers cannot be larger then one. On the contrary,
the spin-$1/2$ double kicked rotor proposed in this work allows topological
phases with arbitrarily large winding numbers to appear. The realization of
ORDKRS should then open the door for experimental explorations
of the interplay between disorder and Floquet topological phases in
large topological invariant regimes, resulting in potentially more
fruitful patterns of Floquet topological Anderson transitions.

\section*{Acknowledgement}
J.G. is supported by the Singapore NRF grant No. NRF-NRFI2017-04 (WBS No. R-144-000-378-281) and the Singapore Ministry of Education Academic Research Fund Tier I (WBS No. R-144-000-353-112).

\appendix
\vspace{0.5cm}

\section{Expression of $\hat{U}_{\ell}$ in position representation}\label{app:Ul}
\setcounter{equation}{0}
\setcounter{figure}{0}
\numberwithin{equation}{section}
\numberwithin{figure}{section}

In this appendix, we expand a bit more on the derivation of $\hat{U}_{\ell}$
($\ell=1,2$) in the two symmetric time frames used in the main text.
In position representation, the Floquet operator $\hat{U}_{\ell}$
in symmetric time frame $\ell$ is written as $\hat{U}_{\ell}=\sum_{\theta}|\theta\rangle\langle\theta|U_{\ell}(\theta)$,
with
\begin{alignat}{1}
U_{1}(\theta)= & e^{-i\frac{{\cal K}_{1}}{2}\sigma_{x}}e^{-i{\cal K}_{2}\sigma_{y}}e^{-i\frac{{\cal K}_{1}}{2}\sigma_{x}},\\
U_{2}(\theta)= & e^{-i\frac{{\cal K}_{2}}{2}\sigma_{y}}e^{-i{\cal K}_{1}\sigma_{x}}e^{-i\frac{{\cal K}_{2}}{2}\sigma_{y}},
\end{alignat}
where ${\cal K}_{1}=K_{1}\cos\theta$ and ${\cal K}_{2}=K_{2}\sin\theta$
as defined in the main text. Using the formula $e^{-i\gamma{\bf n}\cdot\boldsymbol{\sigma}}=\cos(\gamma)-i\sin(\gamma){\bf n}\cdot\boldsymbol{\sigma}$,
with $\boldsymbol{\sigma}=(\sigma_{x},\sigma_{y},\sigma_{z})$ and
${\bf n}$ being a unit vector, we can reorganize $U_{1}(\theta)$
and $U_{2}(\theta)$ as
\begin{alignat}{1}
U_{1}(\theta)= & \cos({\cal K}_{1})\cos({\cal K}_{2})\nonumber \\
-i & [\sin({\cal K}_{1})\cos({\cal K}_{2})\sigma_{x}+\sin({\cal K}_{2})\sigma_{y}],\\
U_{2}(\theta)= & \cos({\cal K}_{1})\cos({\cal K}_{2})\nonumber \\
-i & [\sin({\cal K}_{1})\sigma_{x}+\sin({\cal K}_{2})\cos({\cal K}_{1})\sigma_{y}].
\end{alignat}
With the identifications
\begin{alignat}{1}
\cos(E)= & \cos({\cal K}_{1})\cos({\cal K}_{2}),\\
\sin(E)= & \sqrt{\sin^{2}({\cal K}_{1})\cos^{2}({\cal K}_{2})+\sin^{2}({\cal K}_{2})},\nonumber \\
= & \sqrt{\sin^{2}({\cal K}_{1})+\sin^{2}({\cal K}_{2})\cos^{2}({\cal K}_{1})},
\end{alignat}
where $E$ being the dispersion relation, and
\begin{equation}
n_{1x}=\frac{\sin({\cal K}_{1})\cos({\cal K}_{2})}{\sin(E)},\quad n_{1y}=\frac{\sin({\cal K}_{2})}{\sin(E)},
\end{equation}
\begin{equation}
n_{2x}=\frac{\sin({\cal K}_{1})}{\sin(E)},\quad n_{2y}=\frac{\sin({\cal K}_{2})\cos({\cal K}_{1})}{\sin(E)},
\end{equation}
we can further express $U_{1}(\theta)$ and $U_{2}(\theta)$ as
\begin{alignat}{1}
U_{\ell}(\theta)= & \cos(E)-i\sin(E)(n_{\ell x}\sigma_{x}+n_{\ell y}\sigma_{y})\nonumber \\
= & e^{-iE(\theta)(n_{\ell x}\sigma_{x}+n_{\ell y}\sigma_{y})}\qquad\ell=1,2
\end{alignat}
Finally, identifying the unit vector ${\bf n}_{\ell}=(n_{\ell x},n_{\ell y})$
for $\ell=1,2$, we arrive at the expression $\hat{U}_{\ell}=\sum_{\theta}|\theta\rangle\langle\theta|e^{-iE(\theta){\bf n}_{\ell}\cdot\boldsymbol{\sigma}}$
used in the main text.

\section{Linear growth of winding numbers}\label{app:WNgrowth}
\setcounter{equation}{0}
\setcounter{figure}{0}
\numberwithin{equation}{section}
\numberwithin{figure}{section}
\begin{figure}
	\centering
	\includegraphics[width=.45\textwidth]{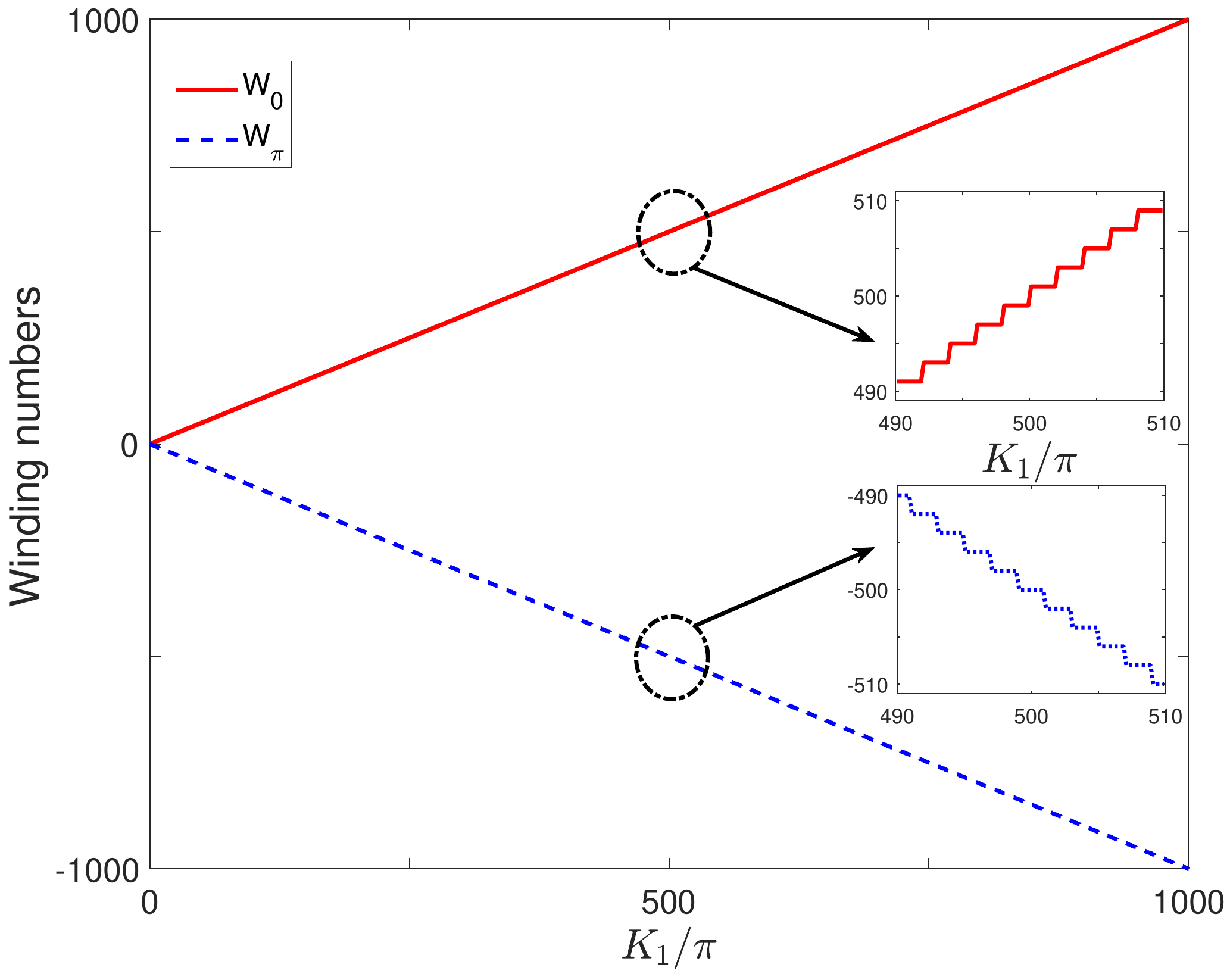}
	\caption{(color online) The linear growth of winding numbers $(W_0,W_{\pi})$ versus kicking strength $K_1$ at a fixed kicking strength $K_2=0.5\pi$.}
	\label{fig:WNgrowth}
\end{figure}
In this appendix, we give an illustration for the change of winding numbers $(W_0,W_{\pi})$ along a trajectory in parallel with the $K_1$-axis at a fixed $K_2\in(0,\pi)$ in the phase diagram Fig.~\ref{fig:PhsDiagram}. From Eq.~(\ref{eq:WN12}), the winding numbers $W_1,W_2$ of Floquet operators $\hat{U}_1,\hat{U}_2$ defined in Eqs.~(\ref{eq:UFrame1},\ref{eq:UFrame2}) are given by
\begin{alignat}{1}
W_{1}=\int_{-\pi}^{\pi}\frac{d\theta}{2\pi}\frac{\sin({\cal K}_{1})\partial_{\theta}{\cal K}_{2}-\sin({\cal K}_{2})\cos({\cal K}_{1})\cos({\cal K}_{2})\partial_{\theta}{\cal K}_{1}}{\sin^{2}(E)},
\end{alignat}
\begin{alignat}{1}
W_{2}=\int_{-\pi}^{\pi}\frac{d\theta}{2\pi}\frac{\sin({\cal K}_{1})\cos({\cal K}_{1})\cos({\cal K}_{2})\partial_{\theta}{\cal K}_{2}-\sin({\cal K}_{2})\partial_{\theta}{\cal
K}_{1}}{\sin^{2}(E)},
\end{alignat}
where $E=\arccos[\cos({\cal K}_{1})\cos({\cal K}_{2})]$,
${\cal K}_{1}=K_{1}\cos\theta$ and ${\cal K}_{2}=K_{2}\sin\theta$.

In our calculation example, we fix $K_2$ at $0.5\pi$ and scan $K_1$ from $0.1\pi$ to $1000\pi$. The results of $\left(W_0=\frac{W_1+W_2}{2},W_{\pi}=\frac{W_1-W_2}{2}\right)$ are presented in Fig.~\ref{fig:WNgrowth}. It is clearly seen that both winding numbers $(W_0,W_{\pi})$ grow linearly with the increase of kicking strength $K_1$.

\section{Calculation of the mean chiral displacement}\label{app:MCD}
\setcounter{equation}{0}
\setcounter{figure}{0}
\numberwithin{equation}{section}
\numberwithin{figure}{section}

In this appendix, we present derivation details of the mean chiral displacement
given by Eq. (\ref{eq:MCD}) of the main text (see also Refs. \cite{MCD1,MCD2}).
For the ORDKRS, A time-frame independent expression of the mean chiral displacement is given by
\begin{equation}
C(t)=\langle0|\otimes\langle F|\hat{U}^{-t}(\hat{n}\otimes\sigma_{z})\hat{U}^{t}|0\rangle\otimes|F\rangle,
\end{equation}
where $|0\rangle$ denotes the $0$-momentum eigenvector and $|F\rangle$
($F=1,2$) denotes the eigenvector of hyperfine level $F$. Note that
for our choice of initial state, $C(t=0)=0$ and $C(t)$ indeed represents
a \emph{displacement} over $t$ driving periods. Writing $\hat{n}$
in momentum representation as $\hat{n}=\sum_{n}n|n\rangle\langle n|$,
we have
\begin{equation}
C(t)=\sum_{n}n\langle0|\otimes\langle F|\hat{U}^{-t}|n\rangle\langle n|\otimes\sigma_{z}\hat{U}^{t}|0\rangle\otimes|F\rangle,
\end{equation}
Expanding $\hat{U}^{t}$ in position representation as $\hat{U}^{t}  = \sum_{\theta}|\theta\rangle\langle\theta|U^{t}(\theta)$,
we further obtain
\begin{alignat}{1}
C(t)= & \sum_{\theta,\theta'}\sum_{n}n\langle0|\theta\rangle\langle\theta'|0\rangle\langle\theta|n\rangle\langle n|\theta'\rangle\nonumber \\
\times & \langle F|U^{-t}(\theta)\sigma_{z}U^{t}(\theta')|F\rangle.
\end{alignat}
Under periodic boundary conditions, we have the following Fourier
transforms between position and momentum basis:
\begin{equation}
|\theta\rangle=\frac{1}{\sqrt{N}}\sum_{n}e^{i\theta n}|n\rangle,\,|n\rangle=\frac{1}{\sqrt{N}}\sum_{\theta}e^{-i\theta n}|\theta\rangle,\,\langle n|\theta\rangle=\frac{1}{\sqrt{N}}e^{i\theta n},
\end{equation}
where $n=-\frac{N}{2},-\frac{N}{2}+1,...,\frac{N}{2}-1$ with $|n\rangle=|n+N\rangle$
and $\theta=\frac{-N\pi}{N},\frac{-(N-1)\pi}{N},...,\frac{(N-1)\pi}{N}$
with $|\theta\rangle=|\theta+2\pi\rangle$. Using these relation,
we can write $C(t)$ as
\begin{equation}
C(t)=\frac{1}{N}\sum_{\theta,\theta'}\frac{1}{N}\sum_{n}ne^{i(\theta'-\theta)n}\langle F|U^{-t}(\theta)\sigma_{z}U^{t}(\theta')|F\rangle.
\end{equation}
Noting that
\begin{equation}
\frac{1}{N}\sum_{n}ne^{i(\theta'-\theta)n}=i\partial_{\theta}\left[\frac{1}{N}\sum_{n}e^{i(\theta'-\theta)n}\right]=i\partial_{\theta}\delta_{\theta\theta'},
\end{equation}
the expression of $C(t)$ reduces to
\begin{equation}
C(t)=\frac{1}{N}\sum_{\theta,\theta'}i\partial_{\theta}\delta_{\theta\theta'}\langle F|U^{-t}(\theta)\sigma_{z}U^{t}(\theta')|F\rangle.
\end{equation}
To proceed, we need to transform the summation over $\theta,\theta'$
to integrals by taking the number of unit cells $N\rightarrow\infty$.
In this limit, we have $\delta_{\theta\theta'}\rightarrow\frac{2\pi}{N}\delta(\theta-\theta')$,
$\sum_{\theta,\theta'}\rightarrow N^{2}\int_{-\pi}^{\pi}\frac{d\theta}{2\pi}\int_{-\pi}^{\pi}\frac{d\theta'}{2\pi}$,
and therefore
\begin{equation}
C(t)=\int_{-\pi}^{\pi}\frac{d\theta}{2\pi}\int_{-\pi}^{\pi}\langle F|U^{-t}(\theta)\sigma_{z}U^{t}(\theta')|F\rangle\left[i\partial_{\theta}\delta(\theta-\theta')\right]d\theta'.
\end{equation}
Sending $i\partial_{\theta}\rightarrow-i\partial_{\theta'}$, performing an integration by parts over $\theta'$ and then integrating out $\theta'$, we are left with
\begin{equation}
C(t)=\int_{-\pi}^{\pi}\frac{d\theta}{2\pi}\langle F|U^{-t}(\theta)\sigma_{z}i\partial_{\theta}U^{t}(\theta)|F\rangle.
\end{equation}
According to our discussion in appendix~\ref{app:Ul}, $U^{t}(\theta)$ can be expressed as
\begin{alignat}{1}
U^{t}(\theta) &= e^{-iE(\theta)t{\bf n}(\theta)\cdot\boldsymbol{\sigma}}\nonumber \\
&= \cos(Et)-i\sin(Et){\bf n}\cdot\boldsymbol{\sigma}=[U^{-t}(\theta)]^{\dagger},
\end{alignat}
where ${\bf n}=(n_{x},n_{y})$ represents the unit vector in any chiral
symmetric time frame, and $\boldsymbol{\sigma}=(\sigma_{x},\sigma_{y})$.
Using this expression of $U^{t}(\theta)$, the operator $U^{-t}(\theta)\sigma_{z}i\partial_{\theta}U^{t}(\theta)$
yields:
\begin{alignat}{1}
& U^{-t}(\theta)\sigma_{z}i\partial_{\theta}U^{t}(\theta)\nonumber \\
= & it\cos(2Et)\left(\partial_{\theta}E\right)(n_{x}\sigma_{y}-n_{y}\sigma_{x})\nonumber \\
+ & i\sin(Et)\cos(Et)\partial_{\theta}(n_{x}\sigma_{y}-n_{y}\sigma_{x})\nonumber \\
- & it\left[\partial_{\theta}\sin^{2}(Et)\right]\sigma_{z}\nonumber \\
+ & \sin^{2}(Et)(n_{x}\partial_{\theta}n_{y}-n_{y}\partial_{\theta}n_{x}).\label{eq:USU}
\end{alignat}
Next, we note that the hyperfine basis $|F=1,2\rangle$ has the following
vector expressions:
\begin{equation}
|F=1\rangle=\begin{pmatrix}1\\
0
\end{pmatrix}\qquad|F=2\rangle=\begin{pmatrix}0\\
1
\end{pmatrix}.
\end{equation}
This means that under the average $\langle F|\cdots|F\rangle$, only
diagonal elements of the matrix $U^{-t}(\theta)\sigma_{z}i\partial_{\theta}U^{t}(\theta)$
could survive. Furthermore, the term $\partial_{\theta}\sin^{2}(Et)$
vanishes after integrating over $\theta$ due to the periodicity of
$E$ in $\theta$. So we are only left with the last term of Eq. (\ref{eq:USU})
under the $\theta$-integral, {\it i.e.},
\begin{equation}
C(t)=\int_{-\pi}^{\pi}\frac{d\theta}{2\pi}\sin^{2}(Et)(n_{x}\partial_{\theta}n_{y}-n_{y}\partial_{\theta}n_{x}).
\end{equation}
Notably, this result is independent of the initial choice of hyperfine
level $|F\rangle$. Finally, with $\sin^{2}(Et)=\frac{1-\cos(2Et)}{2}$,
we arrive at
\begin{equation}
C(t)=\frac{W}{2}-\int_{-\pi}^{\pi}\frac{d\theta}{2\pi}\frac{\cos(2Et)}{2}(n_{x}\partial_{\theta}n_{y}-n_{y}\partial_{\theta}n_{x})
\end{equation}
where $W$ is the winding number in any chiral symmetric time frame,
as given by Eq. (\ref{eq:WN12}) of the main text. More specifically,
for the ORDKRS studied in this work, we have:
\begin{alignat}{1}
C_{1}(t)= & \frac{W}{2}-\int_{-\pi}^{\pi}\frac{d\theta}{2\pi}\frac{\cos(2Et)}{2\sin^{2}(E)}\nonumber \\
\times & \left[\sin({\cal K}_{1})\partial_{\theta}{\cal K}_{2}-\sin({\cal K}_{2})\cos({\cal K}_{1})\cos({\cal K}_{2})\partial_{\theta}{\cal K}_{1}\right],
\\
C_{2}(t)= & \frac{W}{2}-\int_{-\pi}^{\pi}\frac{d\theta}{2\pi}\frac{\cos(2Et)}{2\sin^{2}(E)}\nonumber \\
\times & \left[\sin({\cal K}_{1})\cos({\cal K}_{1})\cos({\cal K}_{2})\partial_{\theta}{\cal K}_{2}-\sin({\cal K}_{2})\partial_{\theta}{\cal K}_{1}\right],
\end{alignat}
where $E=\arccos[\cos({\cal K}_{1})\cos({\cal K}_{2})]$,
${\cal K}_{1}=K_{1}\cos\theta$ and ${\cal K}_{2}=K_{2}\sin\theta$.



\begin{thebibliography}{99}
	
	\bibitem{LeboeufPRL1990} P. Leboeuf, J. Kurchan, M. Feingold, and
	D. P. Arovas, Phys. Rev. Lett. \textbf{65}, 3076 (1990).
	
	\bibitem{KR1} G. Casati, B.V. Chirikov, F.M. Izrailev and J. Ford,
	in \emph{Stochastic Behaviour in classical and Quantum Hamiltonian
		Systems}, Vol. \textbf{93} of Lecture Notes in Physics, edited by
	G. Casati and J. Ford (Springer, New York, 1979).
	
	\bibitem{KR2} G. Casati and B. V. Chirikov, \emph{Quantum Chaos:
		Between Order and Disorder} (Cambridge University Press, New York,
	1995).
	
	\bibitem{KR3} H. Ammann, R. Gray, I. Shvarchuck, and N. Christensen,
	Phys. Rev. Lett. \textbf{80}, 4111 (1998); B. G. Klappauf, W. H. Oskay,
	D. A. Steck, and M. G. Raizen, Phys. Rev. Lett. \textbf{81}, 1203
	(1998).
	
	\bibitem{KR4} Y. Chen and C. Tian, Phys. Rev. Lett. \textbf{113},
	216802 (2014); C. Tian, Y. Chen, and J. Wang, Phys. Rev. B \textbf{93},
	075403 (2016).
	
	\bibitem{KR5} J. Chab\'e, G. Lemari\'e, B. Gr\'emaud, D. Delande, P. Szriftgiser, and J. C. Garreau, Phys. Rev. Lett. {\bf 101}, 255702 (2008).
	
	\bibitem{KR6} D. H. White, S. K. Ruddell, and M. D. Hoogerland, Phys. Rev. A {\bf 88}, 063603 (2013).
	
	\bibitem{OnRes1} C. Ryu, M. F. Andersen, A. Vaziri, M. B. d\textquoteright Arcy,
	J. M. Grossman, K. Helmerson, and W. D. Phillips, Phys. Rev. Lett.
	\textbf{96}, 160403 (2006); I. Talukdar, R. Shrestha, and G. S. Summy,
	Phys. Rev. Lett. \textbf{105}, 054103 (2010).
	
	\bibitem{OnRes2} F. L. Moore, J. C. Robinson, C. F. Bharucha, B.
	Sundaram, and M. G. Raizen, Phys. Rev. Lett. \textbf{75}, 4598 (1995);
	J. F. Kanem, S. Maneshi, M. Partlow, M. Spanner, and A. M. Steinberg,
	Phys. Rev. Lett. \textbf{98}, 083004 (2007); A. Ullah and M. D. Hoogerland,
	Phys. Rev. E \textbf{83}, 046218 (2011).
	
	\bibitem{KRRevs} F. M. Izrailev, Phys. Rep. {\bf 196}, 299 (1990); M. G. Raizen, Advances In Atomic, Molecular, and Optical Physics {\bf 41}, 43 (1999); I. Dana, Can. J. Chem. {\bf 92}, 77 (2014); M. Sadgrovea and S. Wimberger, Advances in Atomic, Molecular, and Optical Physics {\bf 60}, 315 (2011).
	
	\bibitem{DKRGong} J. Wang and J. B. Gong, Phys. Rev. A \textbf{77},
	031405 (2008); J. Wang, A. S. Mouritzen, and J. Gong, J. Mod. Optics
	\textbf{56}, 722 (2009).
	
	\bibitem{Butterfly} D. R. Hofstadter, Phys. Rev. B \textbf{14}, 2239
	(1976).
	
	\bibitem{DerekPRL2012} D. Y. H. Ho and J. Gong, Phys. Rev. Lett.
	\textbf{109}, 010601 (2012).
	
	\bibitem{KHM} T. Geisel, R. Ketzmerick, and G. Petschel, Phys. Rev.
	Lett. \textbf{67}, 3635 (1991); R. Ketzmerick, G. Petschel, and T.
	Geisel, Phys. Rev. Lett. \textbf{69}, 695 (1992).
	
	\bibitem{KHMDana} I. Dana, Phys. Lett. A {\bf 197}, 413 (1995); I. Dana, Phys. Rev. E {\bf 52}, 466 (1995).
	
	\bibitem{HailongPRE2013} H. Wang, D. Y. H. Ho, W. Lawton, J. Wang,
	and J. Gong, Phys. Rev. E \textbf{88}, 052920 (2013).
	
	\bibitem{DerekPRB2014} D. Y. H. Ho and J. Gong, Phys. Rev. B \textbf{90},
	195419 (2014).
	
	\bibitem{OkaPRB2009} T. Oka and H. Aoki, Phys. Rev. B \textbf{79},
	081406 (2009).
	
	\bibitem{LindnerNP2011} N. H. Lindner, G. Refael, and V. Galitski,
	Nat. Phys. \textbf{7}, 490 (2011).
	
	\bibitem{DahlhausPRB2011} J. P. Dahlhaus, J. M. Edge, J. Tworzydlo,
	and C. W. J. Beenakker, Phys. Rev. B \textbf{84}, 115133 (2011).
	
	\bibitem{KitagawaPRB2011}  T. Kitagawa, T. Oka, A. Brataas, L. Fu,
	and E. Demler, Phys. Rev. B \textbf{84}, 235108 (2011).
	
	\bibitem{TongPRB2013} Q.-J. Tong, J.-H. An, J. Gong, H.-G. Luo, and
	C. H. Oh, Phys. Rev. B \textbf{87}, 201109 (2013).
	
	\bibitem{LeonPRL2013} $\acute{{\rm A}}$. G$\acute{{\rm o}}$mez-Le$\acute{{\rm o}}$n
	and G. Platero, Phys. Rev. Lett. \textbf{110}, 200403 (2013).
	
	\bibitem{CayssolRRL2013} J. Cayssol, B. D$\acute{{\rm o}}$ra, F.
	Simon, and R. Moessner, Phys. Status Solidi Rapid Res. Lett. \textbf{7},
	101 (2013).
	
	\bibitem{GrushinPRL2014} A. G. Grushin, $\acute{{\rm A}}$. G$\acute{{\rm o}}$mez-Le$\acute{{\rm o}}$n,
	and T. Neupert, Phys. Rev. Lett. \textbf{112}, 156801 (2014).
	
	\bibitem{Wang2014} R. Wang, B. Wang, R. Shen, L. Sheng, and D. Y.
	Xing, Europhys. Lett. \textbf{105}, 17004 (2014).
	
	\bibitem{ZhouEPJB2014}  L. Zhou, H. Wang, D. Y. H. Ho, and J. Gong,
	Eur. Phys. J. B \textbf{87}, 204 (2014).
	
	\bibitem{TitumPRL2015} P. Titum, N. H. Lindner, M. C. Rechtsman,
	and G. Refael, Phys. Rev. Lett. \textbf{114}, 056801 (2015).
	
	\bibitem{XiongPRB2016} T.-S. Xiong, J. Gong, and J.-H. An, Phys.
	Rev. B \textbf{93}, 184306 (2016).
	
	\bibitem{DLossPRLPRB} J. Klinovaja, P. Stano, and D. Loss, Phys.
	Rev. Lett. \textbf{116}, 176401 (2016); M. Thakurathi, D. Loss, and
	J. Klinovaja, Phys. Rev. B \textbf{95}, 155407 (2017).
	
	\bibitem{KitagawaPRB2010} T. Kitagawa, E. Berg, M. Rudner, and E.
	Demler, Phys. Rev. B \textbf{82}, 235114 (2010).
	
	\bibitem{JiangPRL2011} L. Jiang, T. Kitagawa, J. Alicea, A. R. Akhmerov,
	D. Pekker, G. Refael, J. I. Cirac, E. Demler, M. D. Lukin, and P.
	Zoller, Phys. Rev. Lett. \textbf{106}, 220402 (2011).
	
	\bibitem{KunduPRL2013} A. Kundu and B. Seradjeh, Phys. Rev. Lett.
	\textbf{111}, 136402 (2013).
	
	\bibitem{Bomantara2016} R. W. Bomantara, G. N. Raghava, L. Zhou,
	and J. Gong, Phys. Rev. E \textbf{93}, 022209 (2016); R. W. Bomantara
	and J. Gong, Phys. Rev. B \textbf{94}, 235447 (2016).
	
	\bibitem{ZhaoErH2014} M. Lababidi, I. I. Satija, and E. Zhao, Phys.
	Rev. Lett. \textbf{112}, 026805 (2014); Z. Zhou, I. I. Satija, and
	E. Zhao, Phys. Rev. B \textbf{90}, 205108 (2014).
	
	\bibitem{ReichlPRA2014} M. D. Reichl and E. J. Mueller, Phys. Rev.
	A \textbf{89}, 063628 (2014).
	
	\bibitem{LeonPRB2014} $\acute{{\rm A}}$. G$\acute{{\rm o}}$mez-Le$\acute{{\rm o}}$n,
	P. Delplace, and G. Platero, Phys. Rev. B \textbf{89}, 205408 (2014).
	
	\bibitem{AnomalousESPRX} M. S. Rudner, N. H. Lindner, E. Berg, and
	M. Levin, Phys. Rev. X \textbf{3}, 031005 (2013); P. Titum, E. Berg,
	M. S. Rudner, G. Refael, and N. H. Lindner, Phys. Rev. X \textbf{6},
	021013 (2016).
	
	\bibitem{FulgaPRB2016} I. C. Fulga and M. Maksymenko, Phys. Rev.
	B \textbf{93}, 075405 (2016).
	
	\bibitem{YapPRB2017}  H. H. Yap, L. Zhou, J. Wang, and J. Gong, Phys.
	Rev. B \textbf{96}, 165443 (2017).
	
	\bibitem{YapMajorana2017} H. H. Yap, L. Zhou, C. H. Lee, J. Gong,
	arXiv:1711.09540 (2017).
	
	\bibitem{ZhouPRB2016}  L. Zhou, C. Chen, and J. Gong, Phys. Rev.
	B \textbf{94}, 075443 (2016).
	
	\bibitem{AsbothSTF} J. K. Asb$\acute{{\rm o}}$th, Phys. Rev. B \textbf{86},
	195414 (2012). J. K. Asb$\acute{{\rm o}}$th, and H. Obuse, Phys.
	Rev. B \textbf{88}, 121406 (2013).
	
	\bibitem{NathanNJP2015} F. Nathan and M. S. Rudner, New J. Phys.
	\textbf{17},\textbf{ }125014 (2015).
	
	\bibitem{ClassificationFTP1} R. Roy and F. Harper, Phys. Rev. B \textbf{94},
	125105 (2016); R. Roy and F. Harper, Phys. Rev. B \textbf{96}, 155118
	(2017).
	
	\bibitem{ClassificationFTP2} S. Yao, Z. Yan, and Z. Wang, Phys. Rev.
	B \textbf{96}, 195303 (2017).
	
	\bibitem{EckardtRMP2017} See A. Eckardt, Rev. Mod. Phys. \textbf{89},
	011004 (2017) for a review.
	
	\bibitem{ColdAtomFTP} G. Jotzu, M. Messer, R. Desbuquois, M. Lebrat,
	T. Uehlinger, D. Greif, and T. Esslinger, Nature (London) \textbf{515},
	237 (2014); M. Aidelsburger, M. Lohse, C. Schweizer, M. Atala, J. T.
	Barreiro, S. Nascimb$\grave{{\rm e}}$ne, N. R. Cooper, I. Bloch and
	N. Goldman, Nat. Phys.~\textbf{11}, 162 (2015).
	
	\bibitem{PhotonFTP0} T. Kitagawa, M. A. Broome, A. Fedrizzi, M. S.
	Rudner, E. Berg, I. Kassal, A. Aspuru-Guzik, E. Demler and A. G. White,
	Nat. Commun. \textbf{3}, 882 (2012).
	
	\bibitem{PhotonFTP1} M. C. Rechtsman, J. M. Zeuner, Y. Plotnik, Y.
	Lumer, D. Podolsky, F. Dreisow, S. Nolte, M. Segev, and A. Szameit,
	Nature (London) 496, \textbf{196} (2013); W. Hu, J. C. Pillay, K.
	Wu, M. Pasek, P. P. Shum, and Y. D. Chong, Phys. Rev. X \textbf{5},
	011012 (2015).
	
	\bibitem{PhotonFTP2} L. J. Maczewsky, J. M. Zeuner, S. Nolte, and
	A. Szameit, Nat. Commun. \textbf{8}, 13756 (2017); S. Mukherjee, A.
	Spracklen, M. Valiente, E. Andersson, P. Ohberg, N. Goldman, and R.
	R. Thomson, Nat. Commun. \textbf{8}, 13918 (2017).
	
	\bibitem{PhononFTP} M. Xiao, G. Ma, Z. Yang, P. Sheng, Z. Q. Zhang,
	and C. T. Chan, Nat. Phys. \textbf{11}, 240 (2015); R. S$\ddot{{\rm u}}$sstrunk
	and S. D. Huber, Science \textbf{349}, 47 (2015); R. Fleury, A. B.
	Khanikaev and A. Al$\grave{{\rm u}}$, Nat. Commun. \textbf{7}, 11744
	(2016); R. S$\ddot{{\rm u}}$sstrunk, P. Zimmermann, and S. D. Huber,
	New J. Phys. \textbf{19}, 015013 (2017).
	
	\bibitem{DKREXP} P. H. Jones, M. M. Stocklin, G. Hur, and T. S. Monteiro,
	Phys. Rev. Lett. \textbf{93}, 223002 (2004); C. E. Creffield, G. Hur,
	and T. S. Monteiro, Phys. Rev. Lett. \textbf{96}, 024103 (2006).
	
	\bibitem{BetaEqT0} I. Dana, V. Ramareddy, I. Talukdar, and G. S.
	Summy, Phys. Rev. Lett. \textbf{100}, 024103 (2008); M. Sadgrove,
	M. Horikoshi, T. Sekimura, and K. Nakagawa, Phys. Rev. Lett. \textbf{99},
	043002 (2007).
	
	\bibitem{Wimberger} G. Summy and S. Wimberger, Phys. Rev. A \textbf{93},
	023638 (2016).
	
	\bibitem{Wimberger2} S. Dadras, A. Gresch, C. Groiseau, S. Wimberger,
	G. S. Summy, arXiv:1802.08160 (2018).
	
	\bibitem{KRSOther} Other versions of spin-$1/2$ kicked rotor were also proposed in R. Scharf, J. Phys. A: Math. Gen. {\bf 22}, 4223 (1989); M. Thaha, R. Bl\"umel, and U. Smilansky, Phys. Rev. E {\bf 48}, 1764 (1993); D. R. Masovic, J. Phys. A: Math. Gen. {\bf 28}, L147 (1995); M. Bienert, F. Haug, and W. P. Schleich, Phys. Rev. Lett. {\bf 89}, 050403 (2002); C. Zhang, J. Liu, M. G. Raizen, and Q. Niu, Phys. Rev. Lett. {\bf 92}, 054101 (2004).
	
	\bibitem{MCD1} F. Cardano, A. D\textquoteright Errico, A. Dauphin,
	M. Maffei, B. Piccirillo, C. de Lisio, G. D. Filippis, V. Cataudella,
	E. Santamato, L. Marrucci, M. Lewenstein and P. Massignan, Nat. Commun.
	\textbf{8}, 15516 (2017); M. Maffei, A. Dauphin, F. Cardano, M. Lewenstein
	and P. Massignan, New J. Phys. \textbf{20}, 013023 (2018).
	
	\bibitem{MCD2} E. J. Meier, F. A. An, A. Dauphin, M. Maffei, P. Massignan,
	T. L. Hughes, and B. Gadway, arXiv:1802.02109 (2018).
	
	\bibitem{QWReview} See T. Kitagawa, Quantum Inf. Process {\bf 11}, 1107 (2012) for a review.
	
	\bibitem{Tenfold} S. Ryu, A. P. Schnyder, A. Furusaki, and A. W.
	W. Ludwig, New J. Phys. \textbf{12}, 065010 (2010); C. Chiu, J. C.
	Y. Teo, A. P. Schnyder, and S. Ryu, Rev. Mod. Phys. \textbf{88}, 035005
	(2016).
	
	\bibitem{QHEplateaus}K. v. Klitzing, G. Dorda, and M. Pepper, Phys. Rev. Lett. {\bf 45}, 494 (1980).
	
	
	\bibitem{RadityaMTC}R. W. Bomantara and J. Gong, arXiv:1712.09243 (2018).
	
	\bibitem{TCReview}See K. Sacha and J. Zakrzewski, Rep. Prog. Phys. {\bf 81}, 016401 (2018) for a review.
	
	\bibitem{WNFluct2018} M. Rodriguez-Vega, and B. Seradjeh, arXiv:1706.05303
	(2018).
\end{thebibliography}
\end{document}